\documentclass[sigconf,natbib=false]{acmart}

\usepackage{xcolor}
\usepackage{siunitx} % unit
\DeclareCaptionLabelFormat{andtable}{#1~#2  \&  \tablename~\thetable}

\AtBeginDocument{%
  }

\setcopyright{acmlicensed}
\copyrightyear{2025}
\acmYear{2025}
\acmDOI{XXXXXXX.XXXXXXX}
\acmConference[Conference acronym 'XX]{}{June 03--05,
  2018}{Woodstock, NY}
\acmISBN{978-1-4503-XXXX-X/2018/06}

\newcommand{\apdx}[1]{{\color[rgb]{0.0, 0.0, 0.0}#1}}

\acmSubmissionID{7008}

\RequirePackage[
  datamodel=acmdatamodel,
  style=acmnumeric,
  ]{biblatex}

\begin{document}

\title{CLIO: A Tour Guide Robot with Co-speech Actions for\\ Visual Attention Guidance and Enhanced User Engagement}
% \title{CLIO: A tour guide robot with co-speech actions for enhanced user engagement}


\author{Yuxuan Chen$^*$}
\affiliation{%
  \institution{Tam Wing Fan Innovation Wing,\\ The University of Hong Kong}
      \country{Hong Kong, China}
      }
\email{cyx1024@connect.hku.hk}

\author{Ian Leong Ting Lo$^*$}
\affiliation{%
  \institution{Tam Wing Fan Innovation Wing,\\ The University of Hong Kong}
  \country{Hong Kong, China}
  }
\email{ianltlo@hku.hk}

\author{Bao Guo}
\affiliation{%
  \institution{Tam Wing Fan Innovation Wing,\\ The University of Hong Kong}
  \country{Hong Kong, China}
  }
\email{bguo@hku.hk}

\author{Netitorn Kawmali}
\affiliation{%
  \institution{Tam Wing Fan Innovation Wing,\\ The University of Hong Kong}
  \country{Hong Kong, China}
  }
\email{netitornk@gmail.com}

\author{Chun Kit Chan}
\affiliation{%
  \institution{Tam Wing Fan Innovation Wing,\\ The University of Hong Kong}
  \country{Hong Kong, China}
  }
\email{ryancck@hku.hk}

\author{Ruoyu Wang}
\affiliation{%
  \institution{Tam Wing Fan Innovation Wing,\\ The University of Hong Kong}
  \country{Hong Kong, China}
  }
\email{ruoyu@hku.hk}

\author{Jia Pan}
\affiliation{%
  \institution{School of Computing and Data Science, The University of Hong Kong}
  \country{Hong Kong, China}
  }
\email{jpan@cs.hku.hk}

\author{Lei Yang}
\affiliation{%
  \institution{Tam Wing Fan Innovation Wing,\\ The University of Hong Kong}
  \country{Hong Kong, China}
  }
\email{lyang125@hku.hk}

% \def\thefootnote{*}\footnotetext{These authors contributed equally to this work}


\begin{abstract}
While audio guides can offer rich information about an exhibit, it is challenging for visitors to focus on specific exhibit details based only on the verbal description. 
We present \textit{CLIO}, a tour guide robot with co-speech actions to direct visitors' visual attention and thus enhance the overall user engagement in a guided tour. 
\textit{CLIO} is equipped with designed actions to engage visitors. It builds eye contact with the visitor through tracking a visitor's face and blinking its eyes, or orient their attention by its head movement and laser pointer. We further use a Large Language Model (LLM) to coordinate the designed actions with a given narrative script for exhibition.
We conducted a user study to evaluate the \textit{CLIO} system in a mock-up exhibition of historical photographs. 
We collected feedback from questionnaires and quantitative data from a mobile eye tracker. Experimental results validated that the engaging actions are well designed and demonstrated its efficacy in guiding visual attention of the visitors. It was evidenced that \textit{CLIO} achieved an enhanced engagement compared to the baseline system with only audio guidance.
\end{abstract}





\maketitle

%%%%%%%%%%%%%%%%%%%%%%%%%%%%%%%%%%%%%%%%%%%%%%%
% Introduction
%%%%%%%%%%%%%%%%%%%%%%%%%%%%%%%%%%%%%%%%%%%%%%%

% \begin{figure}
%     \centering
%     \includegraphics[trim=0 0 200 0, clip=true, width=\linewidth]{imgs/teaser_new.png}
%     \caption{\textit{CLIO}, a tour guide robot, is designed to enhance user engagement in museum/exhibition tours through engaging actions, such as eye contact and deictic gestures that are coordinated with the audio narration of the exhibits.}
%     \label{fig:teaser}
%     \vspace{-4mm}
% \end{figure}

\begin{figure}
    \centering
    \includegraphics[width=\linewidth]{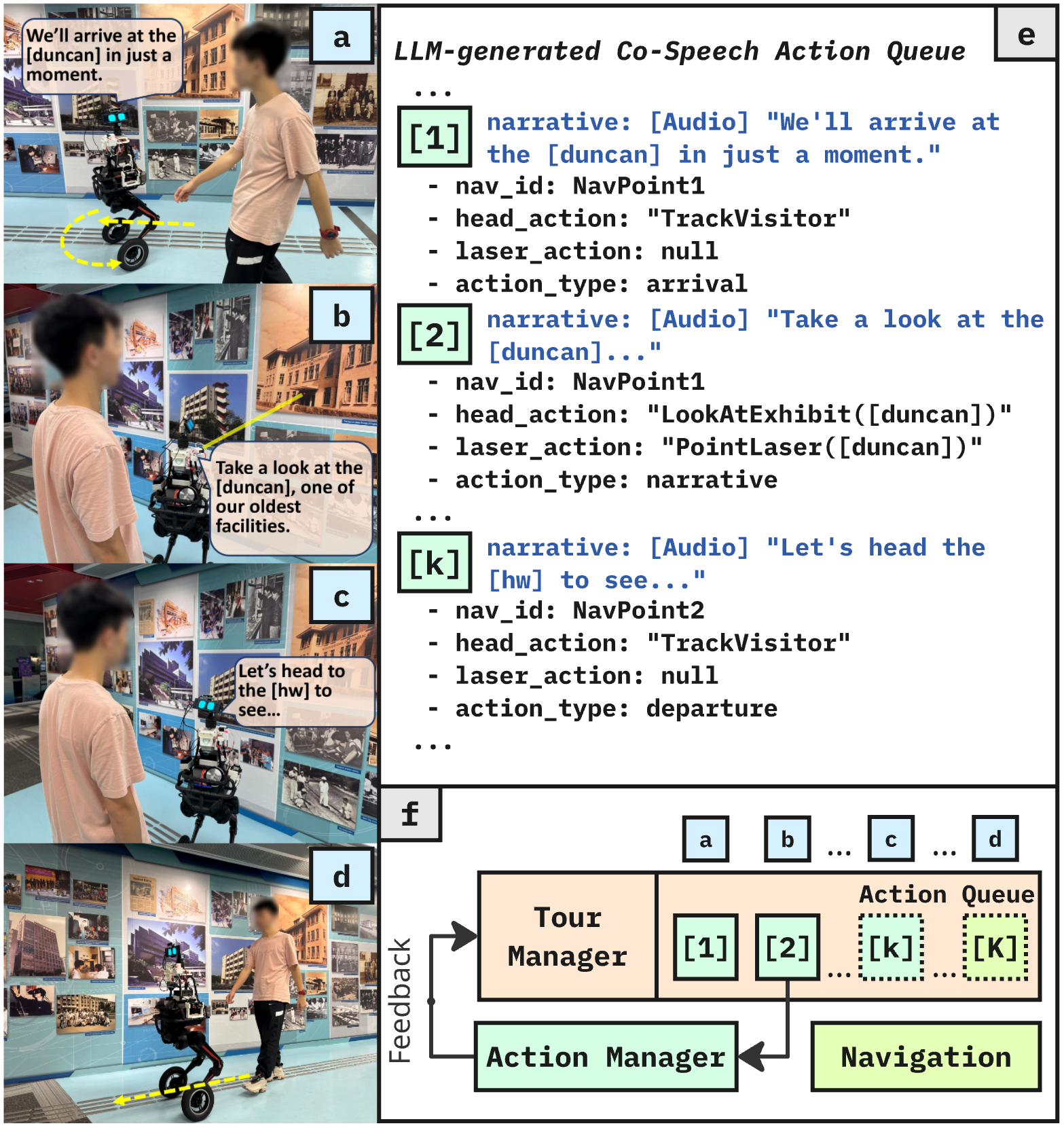}
    \caption{\textit{CLIO}, a tour guide robot, is developed to enhance user engagement in exhibition tours. 
    The system offers coordinated audio-gestural guidance with engaging actions, such as eye contact (a, c) and pointing at the exhibit (b). Our system uses an LLM to generate a queue of actions (e). The tour manager schedules the queue of co-speech actions (solid/dashed border indicates executed or pending actions) (f). The action manager executes concurrent actions, e.g, 2nd section of (e), while the navigation module awaits Action $K$ to perform.
    }
    \vspace{-4mm}
    \label{fig:teaser}
\end{figure}

\section{INTRODUCTION}

While audio guides can offer extensive information, it is challenging for visitors to focus on specific exhibit details based only on the verbal descriptions. In contrast, human docents offer a markedly different experience; for example, they \textit{greet} visitors warmly, \textit{establish eye contact}, and \textit{point} directly to intricate brushwork in a C\'ezanne painting while providing contextual explanations. The absence of \textit{coordinated audio-gestural guidance} significantly disrupts visitors' spatial attention and eventually diminishes engagement. This situation also necessitates increasingly precise narration, yet overly detailed information often fails to align with individual visitors' preferences, creating cognitive overload rather than enhanced understanding. Hence, coordinated audio-gestural behaviors are needed to orient visitors' spatial attention to connect seeing and hearing and enhance visitor engagement in the tour.

Tour guide robots have been studied for decades~\cite{gasteiger2021deploying}. Earlier work focused on fundamental functions of these autonomous systems, such as path planning, localization, and collision avoidance, to deliver functional prototypes. Pioneering systems like RHINO~\cite{burgard1999experiences} at the Deutsches Museum Bonn and MINERVA~\cite{thrun1999minerva} at the Smithsonian's National Museum of American History established the foundational navigation capabilities for museum robotics. 
Building upon these early efforts, recent research has shifted toward more socially-aware designs of robot behaviors~\cite{gehle2017open, bouzida2024carmen, lin2024toward, del2019lindsey, cocsar2020enrichme, mummer} or exploring more effective means (e.g., pointing~\cite{sauppe2014robot, huang2013modeling, holladay2014legible}) other than verbalization for human-robot interaction. We are particularly interested in designing a tour guide robot with socially-aware behaviors to orient visitors' attention and enhance the overall engagement.

We present \textit{CLIO}, a robotic tour guide system as shown in Fig.~\ref{fig:teaser}. \textit{CLIO} is designed to provide coordinated audio-gestural guidance. The system is equipped with a head component consisting of a camera and an LED display. 
This head, mounted with a camera, can rotate in two degrees of freedom (yaw and pitch) to explore the surrounding environment while tracking a visitor's face. 
The display shows a pair of animated eyes that spontaneously blink to build eye contact and engage visitors.
Additionally, a laser pointer is installed to direct visitors' attention to points of interest on the exhibit. These social interaction behaviors are synchronized with each sentence in the audio narration to enhance tour engagement. We will detail the design rationale in Sec.~\ref{sec:methodology}

To validate our design choices, a small-scale user study (Sec.~\ref{sec:experiments}) with 28 participants was conducted in a mock-up exhibition. 
The questionnaire, based on two survey forms~\cite{bartneck2008gqs,obrien2018ues}, investigated the design of robotic actions, the guidance to exhibits, and the user engagement in the robotic guided tour. The subjective results indicate \textit{CLIO} was more acceptable than its Audio-only counterpart and offered more engaging tours with the designed co-speech actions.
The mobile eye tracking data captured during the study evidenced that the co-speech actions offered by \textit{CLIO} effectively directed participants' visual attention as compared to the audio-only condition.

The technical contributions of this work are summarized: 1) A set of actions to guide and engage visitors in an exhibition setup; 2) An integrated system to take a raw script as input and generate a queue of co-speech actions to offer guided tours. 3) A user study and comprehensive analysis of our designed tour guide robot.

%%%%%%%%%%%%%%%%%%%%%%%%%%%%%%%%%%%%%%%%%%%%%%%
% Related Work
%%%%%%%%%%%%%%%%%%%%%%%%%%%%%%%%%%%%%%%%%%%%%%%
\section{RELATED WORK}
    
\textbf{Tour Guide Robots.} 
Building upon the pioneering works of tour guide robots  RHINO~\cite{burgard1999experiences} and MINERVA~\cite{thrun1999minerva}, subsequent works have demonstrated significant advancements in both autonomy and user interaction. 
For instance, Jinny~\cite{kim2004jinny} introduces a lightweight, natural language–based HRI system tailored for untrained users, leveraging keyword extraction and context-aware matching to enable dialogue.
Urbano~\cite{rodriguez2008urbano} presents a comprehensive, full-stack tour-guide robot architecture integrating software and hardware, featuring text-to-speech synthesis, a gesture-capable robotic arm, and an expressive, emotion-conveying facial interface. To further enhance the engagement, Lin et al.~\cite{lin2024toward} introduce a system capable of dynamically adjusting the content according to the user engagement.
Beyond single-robot guided tours, some studies have explored collaborative tour-guiding approaches involving two or more robots working in coordination~\cite{lopez2013guidebot, velentza2019human}.
Interested readers can refer to 
\cite{gasteiger2021deploying} and the references therein for a more holistic picture of robots deployed in museums.

While much of previous work has focused primarily on interaction and engagement, it seems that a crucial yet fundamental characteristic of a competent tour guide-- how to “guide” visitors as a “tour guide” -- has been overlooked.
In this work, we present \textit{CLIO}, a tour guide robot to answer the question: how can a tour guide robot effectively direct visitors’ attention to the exhibits it presents throughout the tour and eventually enhance the tour engagement.

\textbf{Verbal and non-verbal actions for HRI.}
Verbalized communications between human users and robots are important for HRI and have been widely used to facilitate a human-robot team~\cite{hu2019safe, shi2024yell, song2024vlm, unhelkar2020decision}. In a museum setting, Gehle et al.~\shortcite{gehle2017open} studied when and how to greet the visitors to initiate interactions. Meanwhile, non-verbal gestures are widely used in our daily life to communicate and are often perceived as intentional and closely related to particular contexts and situations~\cite{betzler2009expressive, desai2019geppetto, hu2025elegnt, pascher2023communicate}.

Deictic actions are a type of action that orient a person's attention via pointing, which has been proven to be effective in HRI contexts~\cite{holladay2014legible, huang2013modeling, li2023stargazer, sauppe2014robot}.
Besides pointing with a hand-like component, directing the head or body orientation has been explored as a means to direct human attention, for instance, Pereira et al.~\shortcite{pereira2019responsive} discussed how robotic head movement can effectively guide humans' attention. Such head-based cues often serve as foundational elements in establishing joint attention~\cite{anzalone2015evaluating, huang2010joint}, which is regarded as a natural behavior in interaction and enables robots to be more socially interactive. Previous works have primarily investigated the engagement effects of these deictic actions in isolation. Inspired by them, we integrate two deictic actions, i.e., the head-turning action to look at an exhibit and using the laser pointer to project a spot to an exhibit, into a comprehensive tour guide system, to investigate how their combination affects performance in the practical task of museum guiding.

The advent of LLMs provides a low-code interface for generating expressive actions for robots to engage users. Mahadevan et al.~\cite{mahadevan2024generative} proposed an LLM-based approach
to generate expressive actions to engage pedestrians in a daily living set. Compared to~\cite{mahadevan2024generative}, our method is capable of generating a series of coherent and concurrent actions spanning an entire guided tour.

\section{CLIO Tour Guide Robot System}~\label{sec:methodology}

We present \textit{CLIO}, a tour guide robot featuring co-speech guidance actions to engage visitors by orienting their visual attention and thus establishing a better visual-audio connection. Built on top of a wheel-legged robot base with audio output, \textit{CLIO} extends this robot base with two types of engaging actions, i.e., eye contact and deictic gestures via head or laser-pointer movements.

To embody these actions based on a tour script, we propose a tour module that utilizes a Large Language Model (LLM) to parse the script into a queue of actions (Fig.~\ref{fig:teaser}(e)). In each element of the action queue, a narrative sentence is associated with a navigation point and a set of engaging actions. An action manager is developed to synchronize the execution of actions in each element, which then signals the tour manager once all actions are completed (Fig.~\ref{fig:teaser}(f)). We present the design rationale for \textit{CLIO} and then detail the system implementation, including the software, hardware, and the designed co-speech actions for guiding attention and enhancing engagement.

\subsection{Design Rationale}

Some snapshots of the tour are shown in Fig.~\ref{fig:teaser}(a-d)), demonstrating our designed actions in Table~\ref{tab:actions}. 
To establish eye contact with a visitor, our tour guide robot \textit{CLIO} is equipped with a head, namely an LED display showing a pair of animated eyes. \textit{CLIO} uses the head-mounted camera to detect and track the 6D pose of the visitor's face, implemented as the face-tracking action \textsc{TrackVisitor}. 
The \textit{eye-blinking} action, denoted as \textsc{BlinkEye}, is by default activated spontaneously throughout the tour.

Besides establishing eye contact, deictic gestures (e.g., pointing) for directing visitors' attention to specific exhibits are desirable to further enhance visitors' engagement throughout the tour. To this end, we propose two deictic gestures: the \textsc{LookAtExhibit} action, which roughly orients the visitors' attention towards a direction with the head movement, and the \textsc{PointLaser} action that uses a laser pointer to guide the visitors' attention to a specific spot.

Finally, we observed that human docents will manage visitors' expectations by informing them of the upcoming exhibit or moving to another spot. To mimic this practice, we classify a sentence into one of three action types: arrival, narration, and departure. 
 
Despite the availability of engaging actions, programming a robot with rich engaging actions may be tedious based on a given textual script of the tour with a number of areas and exhibits. To overcome this, \textit{CLIO} offers a low-code interface to parse the tour script by leveraging an LLM into an actionable plan that coordinates each narrative description and the designed engaging actions.

\begin{table}[]
    \centering
    \begin{tabular}{l cp{4.5cm}}
        Actions & Type & Description and \{\textcolor{cyan}{\textit{Action parameters}}\}\\
        \hline
        \textsc{PlayAudio} & F & Play \{\textcolor{cyan}{\textit{audio clip}}\} about an exhibit \\
        \hline
        \textsc{BlinkEye} & E-E & Blink the eyes with a \\
        \textsc{TrackVisitor} & E-E & Orient the robot head to \{\textcolor{cyan}{\textit{visitor face}}\}  continuously\\
        \textsc{LookAtExhibit} & E-D & Direct the robot head to \{\textcolor{cyan}{\textit{exhibit}}\} \\
        \textsc{PointLaser} & E-D & 
        Project the laser pointer to highlight \{\textcolor{cyan}{\textit{exhibit}}\} with circular motion patterns \\
        \hline
    \end{tabular}
    \caption{Functional (F) and engaging (E) actions of \textit{CLIO}. We further classify the engaging actions into eye contact type (E-E), deictic type (E-D).}
    \vspace{-4mm}
    \label{tab:actions}
\end{table}

\subsection{System implementation}~\label{sec:system_design}

\textit{CLIO} system consists of five major modules: a tour module, an action manager module, and two functional modules (i.e., a perception module, a navigation module), and a hardware interface module as shown in Fig.~\ref{fig:system}. \textit{CLIO} software system is implemented with Robot Operating System 2 (ROS2)~\cite{doi:10.1126/scirobotics.abm6074} on Ubuntu 22.04. 

\subsubsection{Tour module} 
\begin{figure}[t]
  \centering
  \includegraphics[width=\linewidth]{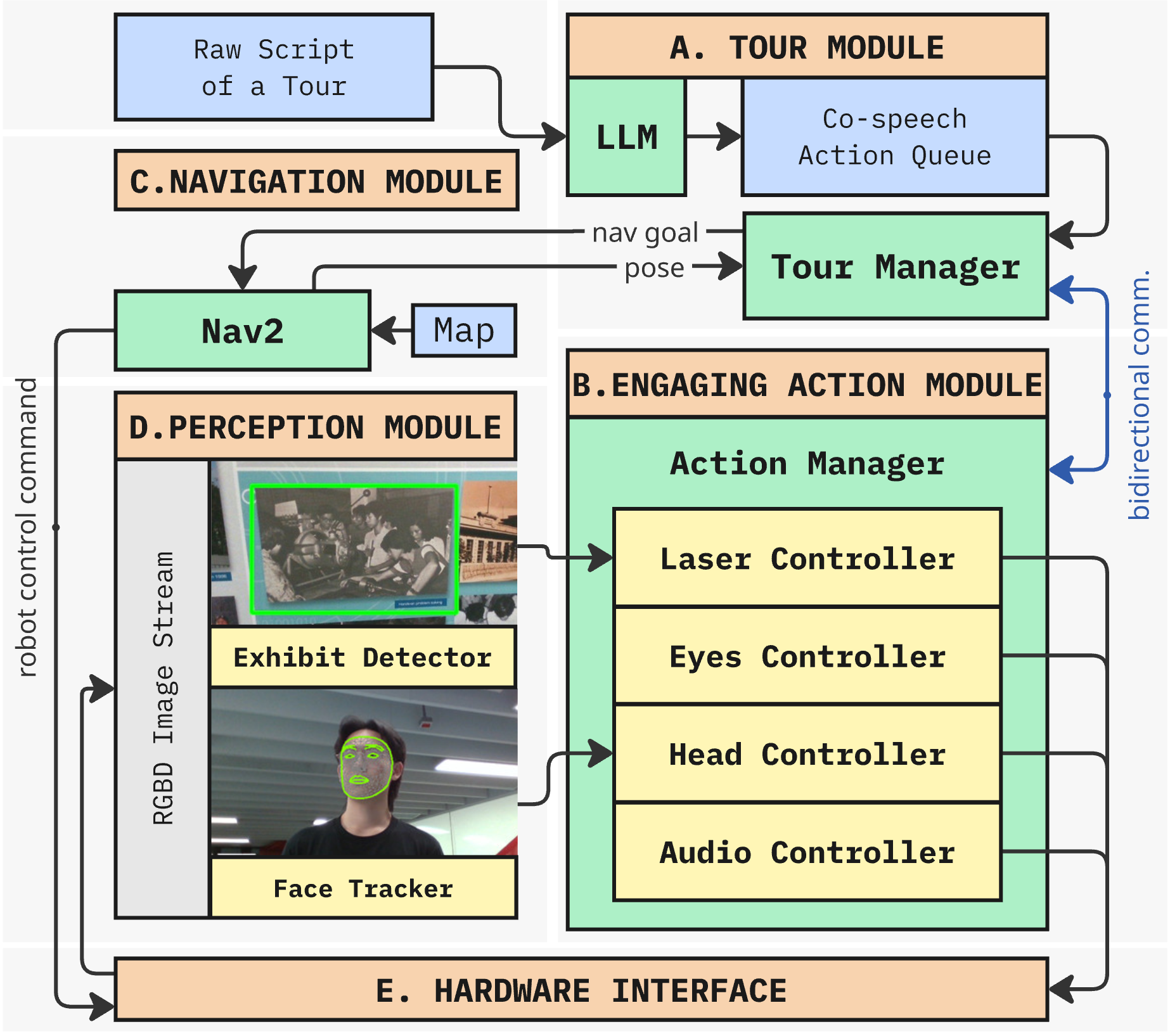}
  \caption{Tour guide system architecture.}
  \label{fig:system}
  \vspace{-4mm}
\end{figure}

As shown in Fig.~\ref{fig:system}(A), \textit{CLIO} system takes a raw script of the tour as input.
An LLM (OpenAI-o3-pro) is used as a low-code interface to parse the script into a \textit{co-speech action queue} to engage visitors.
The LLM takes the script and the descriptions of actions in Table~\ref{tab:actions} as input. 
It is prompted to \textit{segment} the given script into shorter and more engaging sentences while preserving its semantic content, and \textit{associate} each sentence with a set of designed actions. 
For example, when the narrative refers to a particular part of an exhibit (e.g., [duncan]), the LLM will assign both \textsc{LookAtExhibit} and \textsc{PointLaser} actions simultaneously to orient a visitor's visual attention to the exhibit (Fig.~\ref{fig:teaser}(b)).

We also prompt the LLM to extract the navigation point of each exhibit from the given context (i.e., location-exhibit mapping), thereby generating a complete tour plan. Additionally, the LLM is allowed to synthesize transitional narratives that may not exist in the original script to manage the expectations of the visitors.
For example, it can inform visitors en route that they are approaching the next exhibit (Fig.~\ref{fig:teaser}(a)), or notify them before departing to the next location (Fig.~\ref{fig:teaser}(c)).

The action queue output by the LLM is represented as a structured file (Fig.~\ref{fig:teaser}(e)), where each sentence is a handle associated with a set of concurrent actions and the navigation point. \apdx{Due to the limited space, we attach the full LLM prompt and the generated action queues in \textbf{Appendix}.}

During execution, the \textit{online} tour manager will execute the action queue generated by the LLM. 
The robot is modeled as a finite state machine, and the tour manager automatically routes between the state of exhibit introduction processed by the engaging action module or the navigation state by the navigation module.

\subsubsection{Engaging action module}
In this module (Fig.~\ref{fig:system}(B)), an action manager is designed to orchestrate the actions associated with each narrative sentence in the action queue and to perform a timely transition between two sets of actions of consecutive sentences.
To perform the concurrent actions, the action manager must synchronize well the parallel actions within each narrative sentence (termed \textit{intra-sentence} actions); for example, execute the robot to look and point at the exhibit while playing the introduction audio of the exhibit.
Once all intra-sentence actions (including the \textsc{PlayAudio} action) are completed, the robot should seamlessly transit to the next narrative sentence and its associated actions; for example, turning its head back and tracking the visitor's face. This would make the robot's actions more natural.

To implement this, we used the ROS2 action framework, providing built-in progress monitoring and bidirectional communication channels. Each action (e.g., \textsc{PointLaser}, \textsc{TrackVisitor}) is implemented as a ROS2 action server that reports execution status and completion to the action manager. The action manager waits for all actions in the current execution to complete before proceeding to the next sentence. This ensures that speech \textit{only} shifts to the next sentence after all actions (such as \textsc{PointLaser}) have finished, allowing coordinated audio-gestural guidance.

All intra-sentence actions are activated simultaneously along with the audio script in our current implementation. The consideration is based on pre-experiment user feedback suggestions. Otherwise, the action execution may not be considered as timely coordinated. Despite that, our design also permits the possibility of starting intra-sentence actions at different time steps if the action set is further enriched.

Upon receiving the completion notifications from all actions in the previous execution, the action manager signals the tour manager to proceed with the next sentence. This bidirectional communication ensures that actions and speech are properly coordinated throughout the tour. 
It is important to note that \textsc{TrackVisitor}, a continuous action, may span multiple sentences. To prevent abrupt on/off transitions due to sentence changes, the action manager is designed to maintain its execution across consecutive sentences, ensuring a smooth transition between them.

\subsubsection{Functional modules}

\textbf{Perception (Fig.~\ref{fig:system}(D)).}
To engage visitors, a \textit{face tracking} function is implemented with MediaPipe~\cite{lugaresi2019mediapipe} for its efficiency and performance. It takes an RGB image as input and outputs a 3D head model with its 3D position and pose. An example of the detected face is shown in Fig.~\ref{fig:system}.
To localize the exhibits, we trained a YOLOv11 model~\cite{yolo11_ultralytics} as an \textit{exhibit detector} by overfitting it to a set of the same exhibit photographs we collected manually. 
We used the trained detector and the head-mounted RGBD camera to determine the 3D location of each exhibit. The 3D location of an exhibit is computed as the center of all 3D points whose pixels are within the bounding box of the exhibit. The performance of the detector is satisfactory, and no erroneous detection was observed during our experiments.

\textbf{Mapping and navigation (Fig.~\ref{fig:system}(C)).} 
\textit{CLIO} uses FAST-LIO2~\cite{9697912,FASTLIO2_ROS2} to generate a \textit{pre-built map} of a 3D point cloud and use LiDAR-inertial odometry for online localization during run-time. A 2D global occupancy map is obtained from the 3D point cloud for path planning. 
Navigation is achieved by using the Nav2 framework~\cite{macenski2020marathon2}. The NavFn Planner ($A^*$ algorithm) and the MPPI controller~\cite{williams2016aggressive} are adopted for path planning and path tracking while avoiding collision with environmental obstacles. 
Dynamic obstacles (e.g., visitors in the exhibition) are mapped to a $4\times4 m^2$ local occupancy map, which is updated at \SI{10}{Hz} to enable real-time collision avoidance.

To annotate the 3D location of each exhibit in the map, the robot will navigate through the environment while running the exhibit detector combined with depth information to localize and register the position of exhibits.

\subsubsection{Hardware}
\begin{figure}[t]
    \centering
    \includegraphics[width=0.96\linewidth]{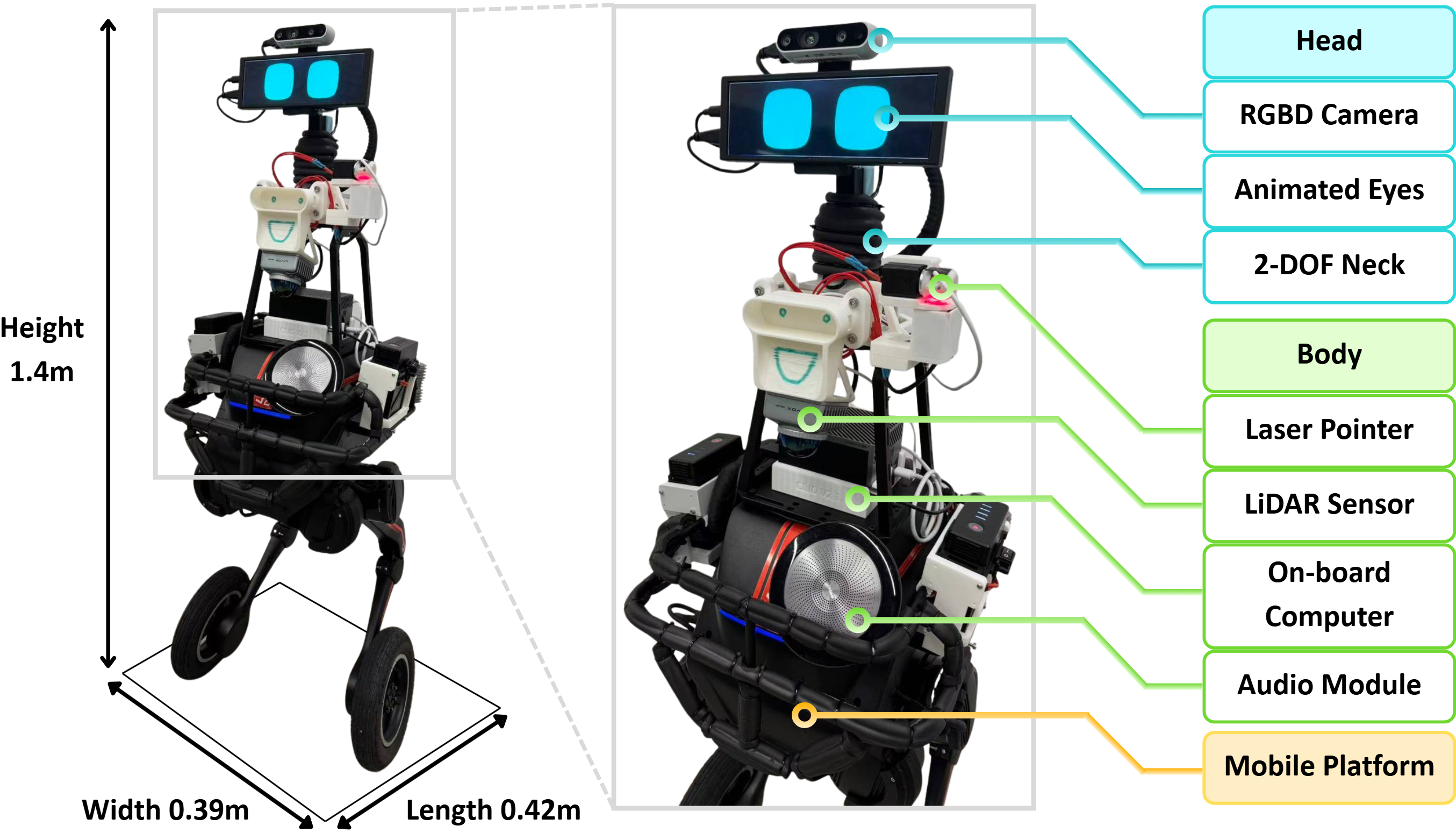}
    \caption{\textit{CLIO} Hardware. The robot is equipped with a head -- an LED screen that displays a pair of animated eyes. An RGB-D camera is mounted on the head.
    The body part houses a laser pointer on its left, a LiDAR sensor, an on-board computer, and an audio speaker. 
    A wheel-legged robot base is adopted to provide an anthropomorphic image to the audience.
    }
    \vspace{-4mm}
    \label{fig:robot_hardware}
\end{figure}

The hardware interface module (Fig.~\ref{fig:system}(E)) controls hardware components of \textit{CLIO} (shown in Figure~\ref{fig:robot_hardware}. 
The \textit{head} consists of an LED display and an RGB-Depth (RGBD) camera (RealSense D435) mounted on top of the display. 
The display shows a pair of animated eyes, as shown in Figure~\ref{fig:robot_hardware}(a), controlled by the eye controller.
Two servo motors, controlled by the Head Controller, are chained to provide two degrees of freedom, namely yaw (rotating left and right) and pitch (rotating up and down).

A \textit{laser pointer} is installed on the left side of the body. The laser pointer gimbal is made of two chained servos controlled by the Laser Controller with yaw and pitch degrees of freedom. 
The laser pointer can cover the entire space of the robot's left side; however, its reachability in the right side of the robot is limited due to laser blockage by the chassis.
To facilitate effective tour guidance, \textit{CLIO} positions itself on the front left of the exhibit and faces the visitor (Fig.~\ref{fig:teaser}(b)), keeping its left side closer to the exhibit to ensure the laser pointer can point to the exhibit without occlusion.

The \textit{body} between the head and the mobile platform houses an on-board computing unit (Nvidia Jetson AGX Orin 64GB), a 3D LiDAR sensor (Livox Mid-360), and an audio speaker (Jabra speaker). A wheel-legged robot is adopted to provide an anthropomorphic image to the audience and is able to navigate in cluttered scenes owing to its small footprint and in-place rotation capability. 

\subsection{Action implementation}

\textbf{\textsc{BlinkEye.}}
To implement natural eye blinking, we followed the recommendations by~\cite{lehmann2016icubblink}, which are based on Doughty's research on human spontaneous blinking during conversation \cite{doughty2001consideration}. 
\textit{CLIO} blinks at an average rate of 23.3 blinks per minute with an inter-eye blink interval of $2.3 \pm \SI{2.0}{s}$ to convey higher perceived intelligence as suggested by \cite{lehmann2016icubblink}.
The blinking has two patterns: single blinks (more common) and occasional double blinks to increase naturalism. By default, the eye blinks spontaneously throughout the tour. We also design the robot to blink on each onset of its verbalization.

\textbf{\textsc{TrackVistor}.}
This \textit{head} action function takes a 3D position to track as input and continuously tracks the given position until overridden by the other head function (i.e., \textsc{LookAtExhibit}).
To efficiently track the visitor, \textit{CLIO} first leverages the difference between the 2D pre-built map and an online-updating 2D local costmap to estimate the visitor's position $(x,y)$ on the floor. 
Then, \textit{CLIO} orients its head display towards the estimated face of the visitor $(x,y,\bar{z}=\SI{1.6}{m})$ and uses the RGBD camera to detect the real visitor's face in 3D using the face detector in the perception module.
Once the visitor's face is successfully detected, the robot tracks the 3D position of the head by solving inverse kinematics of the yaw and pitch angles of the head. The tracking performance can reach an interactive rate ($\sim \SI{10}{Hz}$), allowing the robot to face the visitor with low latency. To stabilize the head movement, we apply a Kalman filter to the detected visitor's face 3D coordinates.  
If no face is detected in the head camera's field of view (FOV) for more than 2 seconds, the head falls back to the default forward-facing direction. Tracking automatically resumes if a face reappears in the camera's FOV.

\textbf{\textsc{LookAtExhibit}.}
The \textit{other} head action provides rough directional guidance for the visitor's visual attention toward the exhibit. It takes the 3D location of an exhibit as input and directs the robotic head toward the exhibit. 3D locations of the exhibits are pre-annotated in the pre-built 3D point cloud map. Meanwhile, this action can also invoke the exhibit detector to ground the exhibit on the map to avoid inaccurate exhibit localization. 

\textbf{\textsc{PointLaser}.}
We equip the robot with a laser pointer to provide detailed visual attention guidance. Given the center of the exhibit in the 3D map, we first extract a patch of point cloud centered around the exhibit from the depth camera. Then, a normal vector is estimated from the sampled point set via singular value decomposition. A circular path on the surface defined by the normal vector is computed from the normal and the center of the exhibit. Given the circular path, the yaw and pitch angles are solved for the laser point's servo, thereby executing the pointing action. The circular motion repeats $k$ times, mimicking human pointing gestures to maintain the visitor's visual attention on the specific exhibit. We empirically set $k=3$.

To precisely point at the exhibit, we need to localize the exhibit during the tour. However, looking at and then pointing at the exhibit will yield a significant latency between the two actions, leading to a seemingly uncoordinated effect.
To handle detection and execution latency, the laser initially circles around the pre-annotated exhibit coordinate, then transitions to the detected location once available.

\section{Hypotheses and Evaluation}~\label{sec:experiments}
\vspace{-4mm}

\subsection{Hypotheses}

We aim to verify the following three hypotheses.
The designed co-speech actions are perceived as natural and intuitive in a tour guide context (\textbf{H1}). 
The designed co-speech actions can guide visitors' visual attention (\textbf{H2}).
The designed co-speech actions can enhance the visitor's engagement during the tour (\textbf{H3}).

\begin{figure}[t]
    \centering
    \includegraphics[width=\linewidth]{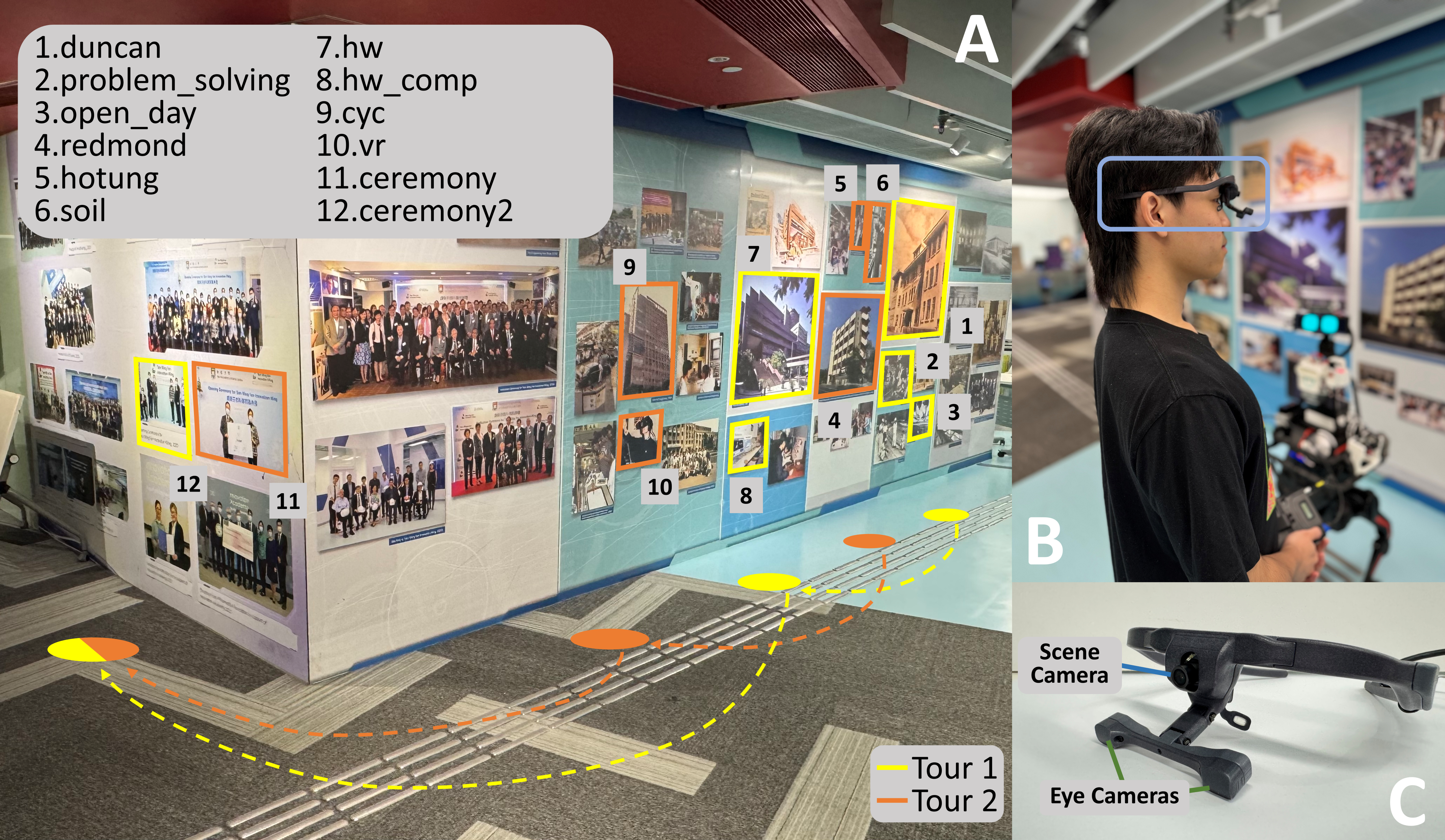}
    \caption{Experimental setup. (A) Two slightly different tour designs. (B) A mobile eye tracker (MET) was used to capture participants' eye gaze during the tour. (C) The MET.
    }
    \vspace{-4mm}
    \label{fig:experiment_setup}
\end{figure}

\subsection{User study design}

To evaluate our hypotheses, we conducted a controlled user study at the photo gallery wall at the [University facility]. 
This user study was approved by the [University IRB] with [approval number].
Each participant experienced two 5-minute guided tours (corresponding to two conditions) provided by the mobile robot shown in Fig.~\ref{fig:robot_hardware}. A mobile eye tracker (MET, Ergoneers Dikablis Glasses 3) was worn by each participant to capture their gaze patterns.

\textbf{Experimental conditions:} 
Each participant experienced the two conditions as follows:
1) \textbf{CLIO}: Participants were offered a guided tour by \textit{CLIO} with all the designed engaging actions listed in Table~\ref{tab:actions}.
2) \textbf{Audio-only}: The robot actions \textsc{TrackVisitor}, \textsc{LookAtExhibit}, and \textsc{PointLaser} were removed compared to the \textit{CLIO} condition. The animated eyes were kept as they are common in market-ready reception robots.

\begin{table}[h]
\centering
\resizebox{0.90\linewidth}{!}{
\begin{tabular}{ccc}
\textbf{Group} & \textbf{First Experience} & \textbf{Second Experience} \\
\hline
A (n=7) & Tour 1 with \textit{CLIO} & Tour 2 with Audio-only \\
B (n=7) & Tour 1 with Audio-only & Tour 2 with \textit{CLIO} \\
C (n=7) & Tour 2 with \textit{CLIO} & Tour 1 with Audio-only \\
D (n=7) & Tour 2 with Audio-only & Tour 1 with \textit{CLIO} \\
\hline
\end{tabular}}
\caption{Counterbalanced experimental design}
\vspace{-4mm}
\label{tab:counterbalancingexpriment}
\end{table}

To prevent learning effects, we developed two distinct tour scripts (Tour 1 and Tour 2) introducing different sets of pictures. 
We implemented a complete counterbalanced design with four groups as shown in Table~\ref{tab:counterbalancingexpriment} to eliminate both order and content effects.
To eliminate the possible confounding factor that different tours provide different experiences to a visitor, we minimized the differences between the two tours as follows.

\textbf{Tour design}: The two tours were designed with similar contents and spatial layouts as shown in Figure \ref{fig:experiment_setup}(A).
Both tours (yellow and orange) share similar navigation paths with three stops to view the pictures. Both tours begin at the right section of the photo wall, then move in a straight line to the second stop two columns away, and conclude at the final stop.
At the first stop, Tour 1 features three pictures, i.e., \textit{duncan} (1), \textit{problem\_solving} (2), and \textit{open\_day} (3), introduced in order. Likewise, Tour 2 introduces \textit{redmond} (4), \textit{hotung} (5), and \textit{soil} (6). In the respective first stops of both tours, \textit{duncan} (1) and \textit{redmond} (6) are the largest of the three pictures. 
The other two pictures are spatially next to each other at a similar height and located either below or above the first picture for both tours. The corresponding script uses directional phrases (e.g., "shown in the picture below," "picture at the top shows") to offer guidance to visitors between pictures. 
The second stop presents \textit{hw} (7) and \textit{hw\_comp} (8) in Tour 1 and \textit{cyc} (9) and \textit{vr} (10) in Tour 2. Both \textit{hw} (7) and \textit{cyc} (9) are located in the middle level with exhibit labels, while \textit{hw\_comp} (8) or \textit{vr} (10) are not physically labelled. Descriptive cues in the photos (i.e., "desktop computer" and "VR headset" are mentioned) are used to inform the visitor about the pictures.  
Both tours conclude at the same stop with \textit{ceremony2} (12) and \textit{ceremony} (11), respectively. We direct attention to a specific photograph by mentioning the number of people it contains. Raw scripts for both tours have nearly identical word counts (Tour 1/2: 292/291 words). \apdx{The raw scripts are provided in Appendix.} Note that in order to attain this high level of similarity between the scripts, the information provided may not be factual. 

\subsubsection{Experimental procedures}
We recruited 28 adult participants (age range: 18-42 years, M = 25.3, SD = 5.42, \#Male = 19, \# Female = 9) from the university community. All participants provided informed consent and had normal or corrected-to-normal vision. None of them had prior experience with the \textit{CLIO} system or a similar one, and they were not familiar with the exhibition content in our setup.

Each participant was assigned to one of the groups (see Table~\ref{tab:counterbalancingexpriment}) to counterbalance the design.
An experimenter first obtained informed consent from a participant and briefed the participant about the experiment procedures. They were specifically told that they would be taking two guided tours offered by a robot to view different pictures and to follow the robot naturally, as they would do in a museum setting. The participant was then fitted with the MET (Fig.~\ref{fig:experiment_setup}(B)). The MET has a scene camera to record the first-person-view video and two eye cameras for each pupil to compute eye gazes (Fig.~\ref{fig:experiment_setup}(C)). A personalized calibration of the MET was performed before each tour. The participant was then advised to position themselves in front of the robot. After the tour, the participant completed a questionnaire followed by a semi-structured interview about their experience. The entire session took $\sim$40 minutes for each participant. After the session, each participant was rewarded with a coupon equivalent to 7 USD.

\subsubsection{Measures} 
\textbf{Questionnaire Measures}: 
We administered a questionnaire after each condition to evaluate the following aspects of the robotic guided tour.
First, \textbf{Robot Impression} (16 items) adopted a part of the Godspeed Questionnaire Series \cite{bartneck2008gqs} (English and local language versions) with three out of the five subscales to measure the participants' perception on the robot: \textit{Animacy} (6 items, measuring how lifelike the robot appeared), \textit{Likeability} (5 items, assessing overall impression), and \textit{Perceived intelligence} (5 items, evaluating competence).
All items used a 5-point semantic differential scale and were presented in randomized order to the participants.
Second, \textbf{Tour Engagement} was measured by \textit{Focused Attention} (7 items) and \textit{Reward Factor} (10 items), a subset of the User Engagement Scale (Long form)~\cite{obrien2018ues}.
These items used a 5-point Likert scale.
Third, we include a customized \textbf{Attention to Exhibits scale} (5 items) to evaluate the self-reported attention of participants toward exhibits. \apdx{The complete questionnaire is attached in the Appendix.}
Following each tour and questionnaire, we conducted 5-10 minute semi-interviews to collect verbal feedback on the tour, 
The interviews began with an open-ended question to capture participants' first impressions of the tour. Subsequent questions probed deeper into the 3 aspects.

\textbf{Eye-tracking Measures.}
We examined the gaze data of each participant and compared the data from the two conditions. Since we are interested in fixations at a particular exhibit with or without our designed actions, we consider a fixation valid if it is within the bounding box of that exhibit. Hence, we computed the following metrics, i.e., the total fixation duration (TFD), average fixation duration (AFD), and time to first fixation (TFF), when the fixation is at a specified exhibit. We additionally report the ratio of TFD over
the total presenting time of an interested exhibit (denoted R-TFD).

\subsection{Results and evaluation}

\subsubsection{Subjective results}

\begin{figure}
    \centering
    \includegraphics[width=\linewidth]{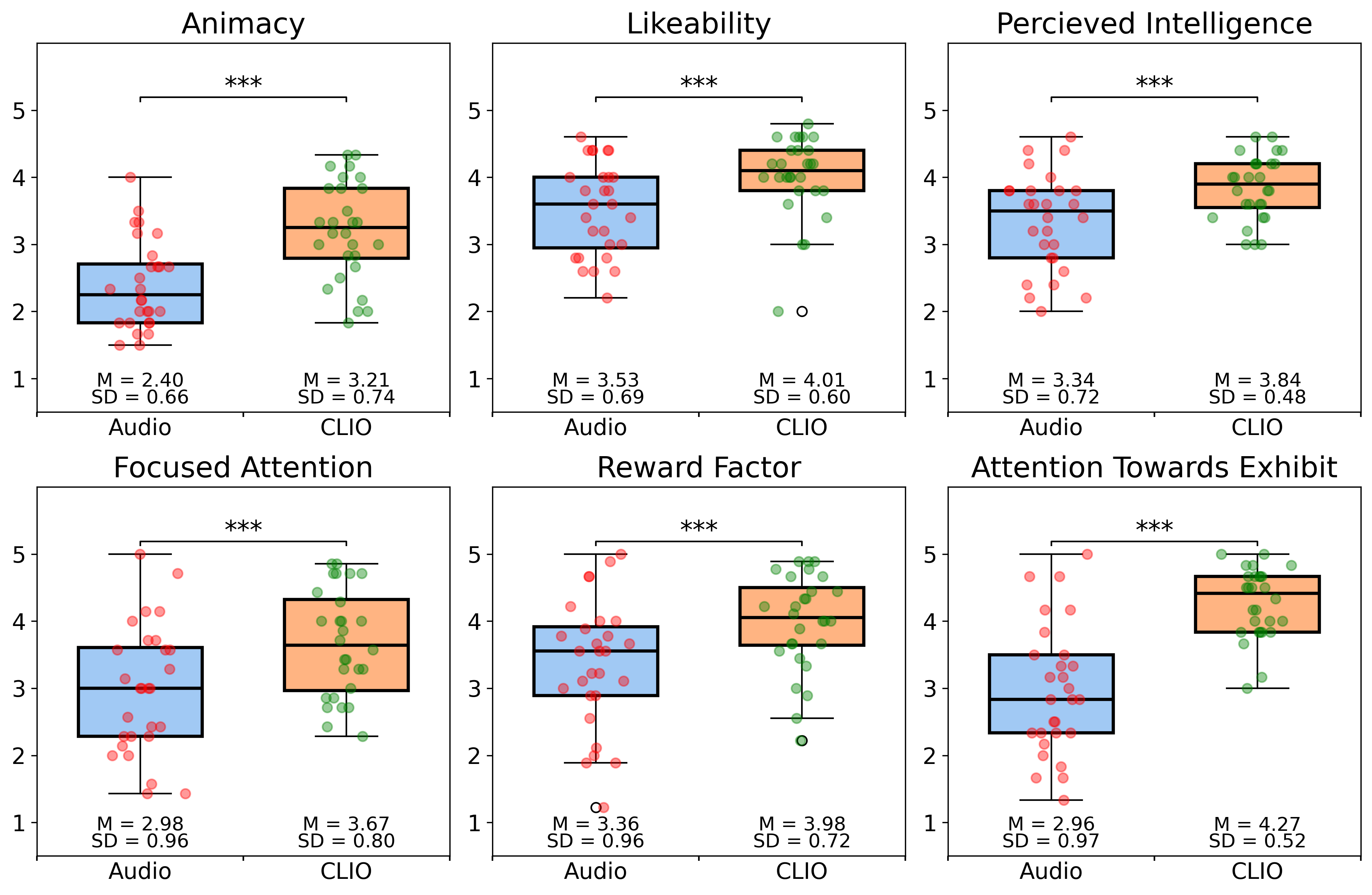}
    \caption{Box plot of questionnaire results. All scales show statistical significance with $p$ < 0.001 (***).}
    \label{fig:questionnaire_results}
    \vspace{-4mm}
\end{figure}

We analyzed the questionnaire data using pairwise t-tests with condition (\textit{CLIO} vs. \textit{Audio-only}) as the within-participants variable. The results are shown in Fig.~\ref{fig:questionnaire_results}. \textit{CLIO} outperformed the \textit{Audio-only} condition on all three dimensions regarding \textbf{Robot Impression}, verifying \textbf{H1} concerning the co-speech action design of \textit{CLIO} (Cronbach’s alpha values are 0.884 for CLIO and 0.934 for Audio-only). A large statistical difference was observed in \textit{Animacy}, where \textit{CLIO} was perceived as more lifelike than the \textit{Audio-only} condition ($p$ < 0.001). 
Statistically significant differences ($p$ < 0.001) were also observed between \textit{CLIO} and the Audio condition regarding both \textit{Likeability} and \textit{Perceived Intelligence}. These findings suggest that the co-speech actions presented by \textit{CLIO} are well-designed and the movements present more intimacy to visitors. 
From the interview, participants noted that the \textit{CLIO}'s actions, such as \textsc{TrackVisitor} and \textsc{LookAtExhibit}, made it feel "alive," "communicative," and "akin to a human tour guide." \textbf{P14} and \textbf{P24} shared similar comments: "I found that the robot is looking at me, and I feel like it is trying to communicate with me," "It makes me feel like a real human is talking to me." They feel these interactive behaviors are similar to a human tour guide directing visitors' attention to exhibits, verifying H1. The robot's actions enhanced visitor engagement, as confirmed by questionnaire ratings, creating a more interactive and lifelike experience.

\textbf{Attention to Exhibits.} The Cronbach's alpha values are 0.75 for \textit{CLIO} and 0.90 for Audio-only for this customized scale.
\textit{CLIO} substantially outperformed the Audio-only condition in directing participants' attention towards exhibits ($p$ < 0.001). This significant difference between the two conditions verifies \textbf{H2}, confirming that the designed actions can effectively guide visitors' visual attention.
From the interview, participants who favored \textit{CLIO} emphasized the laser pointer as a critical feature for precisely locating exhibits. This tool enhanced their attention by guiding them to the exhibit's exact location. From the experiment with Audio-Only mode, \textbf{P25} and \textbf{P6} shared their struggle "I don't know what the robot is doing at the beginning, and it did not indicate where I should look at," "It is very challenging to follow the tour only by hearing, just like a listening exam". 
The PointerLaser supported \textbf{H2} by reducing cognitive load, helping visitors follow the tour flow, and maintaining focus in a visually complex environment. The combination of LookAtExhibit and PointerLaser has further enhanced the effect. \textbf{P19} and \textbf{P28} can locate the laser easier with the head motion "(The head) helps me to find the precise location of the laser," "(The Head) notifies me when to look at the wall". The combination of actions has further verified \textbf{H2} by drawing more attention from the visitor to the tour.

\begin{table}[]
    \centering
    \resizebox{0.45\textwidth}{!}{
    % \begin{tabular}{|c|rr|rr|rr|c|rr|rr|rr|}
    \begin{tabular}{c|rr|rr|rr}
    \hline
    Tour 1 & \multicolumn{2}{c|}{TFF $\downarrow$ } & \multicolumn{2}{c|}{TFD (Ratio) $\uparrow$} & \multicolumn{2}{c}{AFD $\uparrow$} \\
                     & CLIO & Audio & CLIO & Audio & CLIO & Audio \\ 
    \hline
    duncan           & 2.02 & 6.92 & 8.05(0.47) & 2.26(0.20) & 1.15 & 0.66 \\ 
    problem\_solving & 1.09 & 7.30 & 4.25(0.30) & 0.65(0.16) & 0.49 & 0.09 \\ 
    open\_day        & 2.72 & 6.73 & 2.53(0.21) & 0.46(0.04) & 0.25 & 0.17 \\
    hw               & 0.47 & 7.91 & 9.11(0.53) & 2.81(0.18) & 1.21 & 0.63 \\
    hw\_comp         & 1.58 & 5.25 & 4.05(0.29) & 1.87(0.16) & 0.35 & 0.38 \\ 
    ceremony         & 2.04 & 9.92 & 6.82(0.32) & 1.28(0.07) & 1.01 & 0.26 \\
    \midrule
    Tour 2 & \multicolumn{2}{c|}{TFF $\downarrow$ } & \multicolumn{2}{c|}{TFD (Ratio) $\uparrow$} & \multicolumn{2}{c}{AFD $\uparrow$} \\
              & CLIO & Audio & CLIO & Audio & CLIO & Audio \\
    \hline
    redmond   & 1.85 & 9.16 & 9.43(0.43) & 2.63(0.16) & 0.82 & 0.34 \\
    hotung    & 2.01 & 7.24 & 2.14(0.18) & 0.09(0.01) & 0.30 & 0.01 \\
    soil      & 1.64 & 6.55 & 3.53(0.28) & 0.54(0.04) & 0.41 & 0.06 \\
    cyc       & 1.33 & 2.84 & 7.45(0.36) & 3.98(0.21) & 1.06 & 0.45 \\
    vr        & 3.34 & 5.09 & 1.99(0.17) & 1.11(0.10) & 0.25 & 0.27 \\
    ceremony2 & 3.18 & 4.54 & 6.21(0.36) & 3.16(0.21) & 0.67 & 0.64 \\
    \hline
    \end{tabular}}
    \caption{Eye tracking data. Four metrics were calculated for the interested exhibit with \textit{CLIO} and Audio-only, TFF,
    TFD, and AFD. The ratio in the parentheses is TFD over the total presenting time of an interested exhibit. Unit: Seconds}
    \label{tab:eye_tracking_quantitative}
    \vspace{-4mm}
\end{table}

\textbf{Tour Engagement.} Our results verify \textbf{H3} concerning the engagement during the tour (Cronbach’s alpha values are 0.942 for \textit{CLIO} and 0.986 for Audio-only). \textit{CLIO} shows a higher level of \textit{Focused Attention} than that of the \textit{Audio-only} condition ($p$ < 0.001). 
Similarly, \textit{CLIO} also received a higher \textit{Reward Factor} rating on average, suggesting that participants found the tour with co-speech actions more engaging ($p$ < 0.001). 
This shows that the proposed actions in \textit{CLIO} contribute greatly to visitor engagement, achieving beyond what narrative content alone can do.
From the interview, some participants described the robot-guided experience as "Immersive" and "Interactive". \textbf{P16} and \textbf{P26} found the experience unexpected "(\textit{CLIO} guided me smoothly to exhibit), and I can fully focus to the content and I will say it is quite immersive,"  "I thought robot is mechanical but this robot is surprisingly more interactive than my expectation". These comments attributed these qualities to the robot’s engaging actions, which verified H3. The robot's ability to direct their attention, using concurrent actions like the laser pointing and the turning head, enhanced engagement and created a more compelling tour experience.

\subsubsection{Quantitative results on the visual attention}

\begin{figure}[t]
    \centering
    \includegraphics[width=\linewidth]{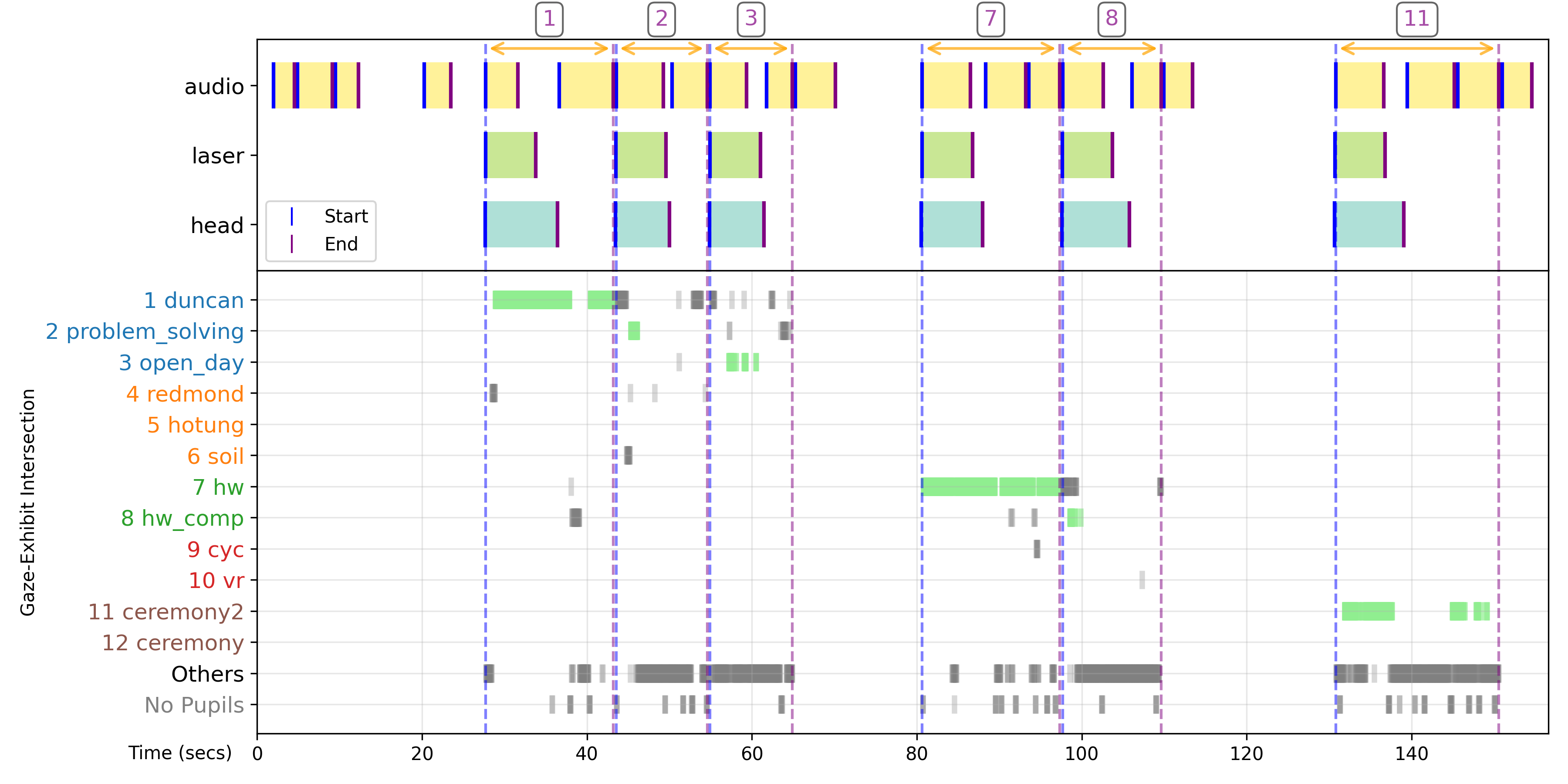}
    \includegraphics[width=\linewidth]{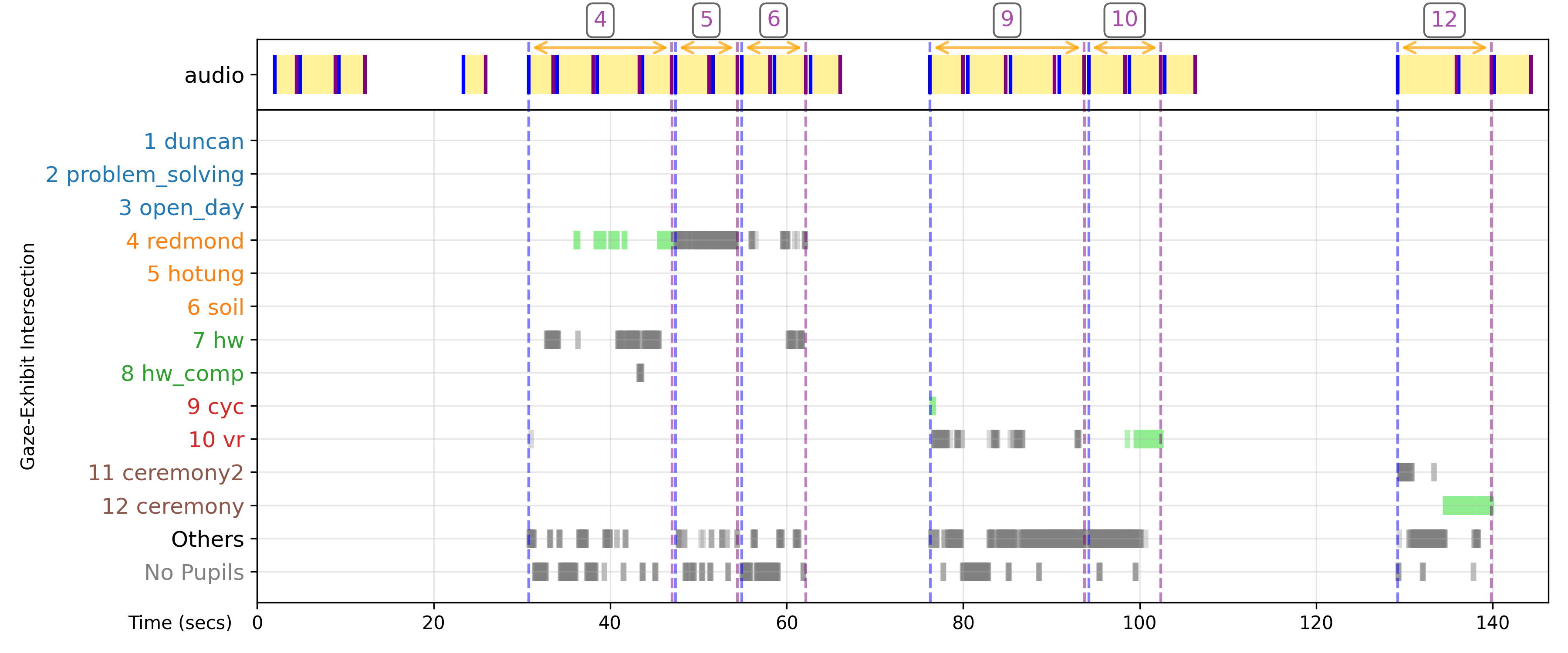}
    \caption{Eye tracking data of Subject id: 23 (group: A). Top: \textit{CLIO}, Tour 1. Bottom: Audio-only, Tour 2}
    \vspace{-4mm}
    \label{fig:eye_tracking_quantitative}
\end{figure}

\begin{figure*}[t]
    \centering
    \includegraphics[width=0.96\linewidth]{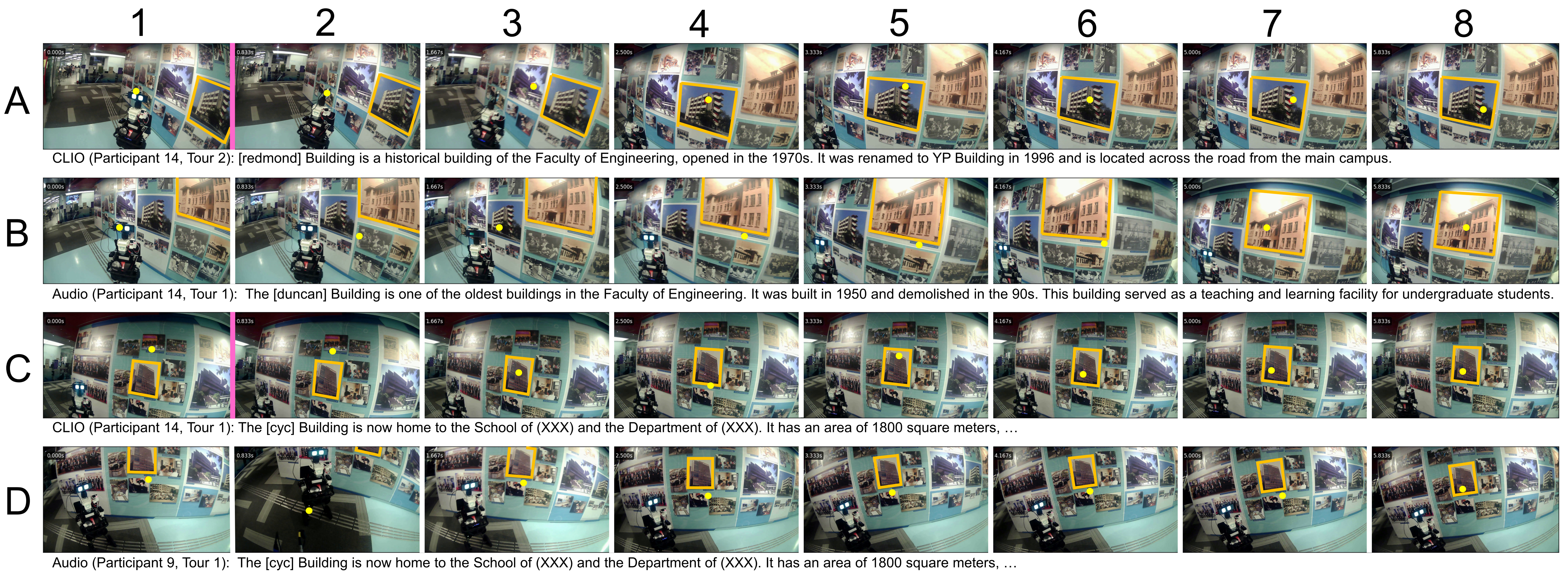}
    \vspace{-4mm}
    \caption{Two pairs of video segments are visualized. See the main text for details.
    }
    \label{fig:paraller_tracker}
\end{figure*}

As presented in Table~\ref{tab:eye_tracking_quantitative}, the TFF for the \textit{CLIO} tour appears consistently lower than that for the Audio tour, indicating that participants in the \textit{CLIO} tour were able to locate the target exhibit more quickly. This time difference in TTF is particularly evident for exhibits that lack distinctive visual features or share similar content. To elaborate, when introducing pictures of buildings, such as \textit{duncan} (1), \textit{redmond} (4), \text{hotung} (5), and \textit{hw} (7), without referring to the descriptive labels beneath each exhibit, participants struggled to distinguish between them based on features of the buildings alone, as the structures show high visual similarity. Consequently, the average TFF value of each exhibit exceeds 6 seconds (\textit{duncan}: 6.92s, \textit{redmond}: 9.16s, \text{hotung}: 7.24s, \textit{hw}: 7.91s). In contrast, under the \textit{CLIO} tour, the robot’s deictic gestures effectively disambiguated the exhibits, leading to remarkable reductions in TFF across all four exhibits (\textit{duncan}: 2.02s, \textit{redmond}: 1.85s, \text{hotung}: 2.01s, \textit{hw}: 0.47s).
Furthermore, most of TFD and AFD from the \textit{CLIO} tour are higher compared to the Audio tour, except for \textit{hw\_comp} (8) and \textit{vr} (10). This suggests that participants were more visually engaged with the exhibits introduced in \textit{CLIO} tour. 

Fig.~\ref{fig:eye_tracking_quantitative} illustrates the eye-tracking data for subject 23, spanning from the beginning to the end of the tour. The top and bottom graphs, respectively, present results from \textit{CLIO} and the Audio-only condition. In each graph, the upper section depicts the robot actions (referred as the Action section) and the lower section the Gaze-Exhibit Intersections (referred as the Gaze section). The $x$-axis represents the time.
The Action section illustrates the duration of each individual action, with the above ID indicating the specific exhibit the robot presented and the corresponding time span. The Gaze section depicts the subject's gaze throughout the tour. Green gazes highlight the subject gazing at the target exhibits. Otherwise, gray is applied. Blue/purple dashed lines mark the start/end of the exhibit presentation by the robot.

The data presented in Fig. \ref{fig:eye_tracking_quantitative} aligns closely with the information in Table \ref{tab:eye_tracking_quantitative}. Notably, the initial occurrence of green highlights under the \textit{CLIO} condition appears sooner than in the audio-only condition (as indicated by TFF in the table). Additionally, green highlights are not only more frequent but also take up a larger percentage of the total presentation time for each exhibit.

Fig.~\ref{fig:paraller_tracker} presents representative scene-camera video segments comparing participant gaze behavior across conditions. The first pair (A and B) shows the same participant in different conditions. Note that different exhibits were presented in different tours. Pairs C and D present the same tour between different subjects and conditions. With \textit{CLIO}'s deictic actions (A and C), participants' gaze (yellow dot) quickly arrived at the exhibit of interest (orange bounding boxes).  In the \textit{Audio-only} condition (B and D), the gaze moved around to check the exhibit label (B4-6, D5-7) before arriving at the exhibit of interest. Pink vertical lines mark the start of the deictic actions.
This comparison evidences that the \textit{CLIO} tour, augmented with \textsc{LookAtExhibit} and \textsc{PointLaser}, helps participants locate the target exhibit quickly. In the Audio tour, by contrast, visitors have to search for the descriptive labels beneath the exhibits to infer which one is being discussed. This observation is consistent with the quantitative eye-tracking results presented in Table~\ref{tab:eye_tracking_quantitative}.
\vspace{-2mm}

\subsubsection{Validation of other key designs}

\textbf{Variations in the generated action queues. }
We conducted multiple test runs with the \textit{OpenAI o3-pro} model to evaluate the consistency of LLM-generated co-speech action queues. We observed natural paraphrasing variations in the narrative content. Moreover, structural patterns of action sequences are very consistent. Multiple test runs output a similar action pattern: (1) \textsc{TrackVisitor} during arrivals and departures; (2) \textsc{LookAtExhibit([Exhibit])} shortly for the first one to two sentences of narrative; (3) \textsc{TrackVisitor} again during the narration until the departure. 

\textbf{Sanity check for LLM-generated plans:} For robustness, a sanity check is conducted to ensure all required input parameters of an action are present. For example, the \textsc{PointLaser} action cannot function without valid exhibit coordinates. In our implementation, the online tour manager maintains a list of exhibit coordinates within the map during a tour. If there are missing exhibit coordinates, the tour manager inserts a \textsc{LookAtExhibit} action to turn the head to the pre-annotated exhibit coordinate, re-localizing the exhibit around its associated navigation point. Empirically, this simple check works well for our experimental setup and was employed to ground the exhibits into our scene map. 
Future work shall consider more advanced methods (e.g.,~\cite{aeronautiques1998pddl, chen2024language,tantakoun2025llms}) to handle complicated scenes with more exhibits or complex robot actions.

\subsection{Discussion}

\textbf{Negative feedback from participants.} 
Some noted that the \textsc{TrackVisitor} action occasionally shifted their attention toward the robot’s head rather than the exhibit, disturbing their engagement with the content. These comments aligned with observations made by previous work~\cite{bitgood2010attention} where stimuli may be distractors for part of the users, demonstrating the need for future study in how to design personalized, engaging actions.

\textbf{User interaction.} Two important aspects should be considered by future studies. First, while \textit{CLIO} can engage the visitors, the visitors cannot interact with the robot directly. A speech interface and a built-in AI-empowered chatbot may be useful if the latency is reduced to an acceptable range. Second, an interesting future work is to conduct an interview to learn museum/exhibition curators' comments on \textit{CLIO} for further development. 

\textbf{Group visitors. }
Our experiment involved only a single participant each time, which is different from a guided tour for group visitors. In this case, the robot is expected to engage multiple visitors, which is an interesting future direction. 

\textbf{Spatial positioning. } 
Another interesting and unexplored question is where the robot should stand between the exhibits and the visitors, to allow effective guidance and avoid occluding the visitors' eyesight. Now, we set the robot to the left of the exhibit to allow its laser pointer to project a spot on the exhibit. Spatial understanding should be incorporated to improve the current system.

\section{Conclusion}

In this paper, we present our robotic system \textit{CLIO} that enables coordinated audio-gestural guidance in an exhibition tour, featuring a set of actions to engage visitors in an exhibition set-up. It takes a tour script as input and generates a co-speech action queue to accompany the narrative. Through a user study on 28 participants, we demonstrate that the designed actions were natural and lifelike, and successfully guided the visual attention of visitors on a tour. Enhanced engagement of the visitors was observed from both eye-tracking data and the questionnaire feedback between our \textit{CLIO} and the Audio-only baseline.


\printbibliography

@article{cocsar2020enrichme,
  title={ENRICHME: Perception and Interaction of an Assistive Robot for the Elderly at Home},
  author={Co{\c{s}}ar, Serhan and Fernandez-Carmona, Manuel and Agrigoroaie, Roxana and Pages, Jordi and Ferland, Fran{\c{c}}ois and Zhao, Feng and Yue, Shigang and Bellotto, Nicola and Tapus, Adriana},
  journal={International Journal of Social Robotics},
  volume={12},
  number={3},
  pages={779--805},
  year={2020},
  publisher={Springer}
}

@artical{mummer,
author = {Foster, Mary and Craenen, Bart and Deshmukh, Amol and Lemon, Oliver and Bastianelli, Emanuele and Dondrup, Christian and Papaioannou, Ioannis and Vanzo, Andrea and Odobez, Jean-Marc and Canévet, Olivier and Cao, Yuanzhouhan and He, Weipeng and Martínez-González, Angel and Motlicek, Petr and Siegfried, Rémy and Alami, Rachid and Belhassein, Kathleen and Buisan, Guilhem and Clodic, Aurélie and Tammela, Antti},
year = {2019},
month = {09},
pages = {},
title = {MuMMER: Socially Intelligent Human-Robot Interaction in Public Spaces},
doi = {10.48550/arXiv.1909.06749}
}

@inproceedings{lin2024toward,
  title={Toward personalized tour-guide robot: Adaptive content planner based on visitor's engagement},
  author={Lin, Yanran and Jo, Wonse and Ali, Arsha and Robert Jr, Lionel P and Tilbury, Dawn M},
  booktitle={Companion of the 2024 ACM/IEEE International Conference on Human-Robot Interaction},
  pages={674--678},
  year={2024}
}

@INPROCEEDINGS{velentza2019human,
  author={Velentza, Anna-Maria and Heinke, Dietmar and Wyatt, Jeremy},
  booktitle={2019 28th IEEE International Conference on Robot and Human Interactive Communication (RO-MAN)}, 
  title={Human Interaction and Improving Knowledge through Collaborative Tour Guide Robots}, 
  year={2019},
  volume={},
  number={},
  pages={1-7},
  doi={10.1109/RO-MAN46459.2019.8956372}
}

@article{lopez2013guidebot,
author = {López, Joaquin and Pérez Losada, Diego and Santos, María and Diaz-Cacho, Miguel},
year = {2013},
month = {01},
pages = {},
title = {GuideBot. A Tour Guide System Based on Mobile Robots},
volume = {10},
journal = {International Journal of Advanced Robotic Systems},
doi = {10.5772/56901}
}

@inbook{rodriguez2008urbano,
author = {Rodríguez-Losada, Diego and Matia, Fernando and Galan, Ramon and Hernando, Miguel and Montero, Juan and Lucas, Juan},
year = {2008},
month = {07},
pages = {},
title = {Urbano, an Interactive Mobile Tour-Guide Robot},
isbn = {978-953-7619-02-2},
journal = {Advances in Service Robotics},
doi = {10.5772/5950}
}

@inproceedings{kim2004jinny,
author = {Kim, Gunhee and Chung, Woojin and Kim, Kyung-Rock and Kim, Munsang and Han, Sangmok and Shinn, R.H.},
year = {2004},
month = {01},
pages = {3450 - 3455 vol.4},
title = {The autonomous tour-guide robot Jinny},
volume = {4},
isbn = {0-7803-8463-6},
journal = {2004 IEEE/RSJ International Conference on Intelligent Robots and Systems (IROS)},
doi = {10.1109/IROS.2004.1389950}
}

@inproceedings{mahadevan2024generative,
  title={Generative expressive robot behaviors using large language models},
  author={Mahadevan, Karthik and Chien, Jonathan and Brown, Noah and Xu, Zhuo and Parada, Carolina and Xia, Fei and Zeng, Andy and Takayama, Leila and Sadigh, Dorsa},
  booktitle={Proceedings of the 2024 ACM/IEEE International Conference on Human-Robot Interaction},
  pages={482--491},
  year={2024}
}

@article{betzler2009expressive,
  title={Expressive actions},
  author={Betzler, Monika},
  journal={Inquiry},
  volume={52},
  number={3},
  pages={272--292},
  year={2009},
  publisher={Taylor \& Francis}
}

@article{doi:10.1126/scirobotics.abm6074,
    author = {Steven Macenski and Tully Foote and Brian Gerkey and Chris Lalancette and William Woodall},
    title = {Robot Operating System 2: Design, architecture, and uses in the wild},
    journal = {Science Robotics},
    volume = {7},
    number = {66},
    pages = {eabm6074},
    year = {2022},
    doi = {10.1126/scirobotics.abm6074},
    URL = {https://www.science.org/doi/abs/10.1126/scirobotics.abm6074}
}

@ARTICLE{9697912,
  author={Xu, Wei and Cai, Yixi and He, Dongjiao and Lin, Jiarong and Zhang, Fu},
  journal={IEEE Transactions on Robotics}, 
  title={FAST-LIO2: Fast Direct LiDAR-Inertial Odometry}, 
  year={2022},
  volume={38},
  number={4},
  pages={2053-2073},
  doi={10.1109/TRO.2022.3141876}}

@software{FASTLIO2_ROS2,
  author = {Lianghe Ming},
  title = {FASTLIO2\_ROS2},
  year = {2024},
  url = {https://github.com/liangheming/FASTLIO2_ROS2},
  note = {GitHub repository},
  version = {latest} % 可选，写具体commit或release
}

@InProceedings{macenski2020marathon2,
author = {Macenski, Steven and Martin, Francisco and White, Ruffin and Ginés Clavero, Jonatan},
title = {The Marathon 2: A Navigation System},
booktitle = {2020 IEEE/RSJ International Conference on Intelligent Robots and Systems (IROS)},
year = {2020}
}

@article{hu2025elegnt,
  title={ELEGNT: Expressive and Functional Movement Design for Non-anthropomorphic Robot},
  author={Hu, Yuhan and Huang, Peide and Sivapurapu, Mouli and Zhang, Jian},
  journal={arXiv preprint arXiv:2501.12493},
  year={2025}
}

@software{yolo11_ultralytics,
  author = {Glenn Jocher and Jing Qiu},
  title = {Ultralytics YOLO11},
  version = {11.0.0},
  year = {2024},
  url = {https://github.com/ultralytics/ultralytics},
  orcid = {0000-0001-5950-6979, 0000-0003-3783-7069},
  license = {AGPL-3.0}
}

@inproceedings{thrun1999minerva,
  title={MINERVA: A second-generation museum tour-guide robot},
  author={Thrun, Sebastian and Bennewitz, Maren and Burgard, Wolfram and Cremers, Armin B and Dellaert, Frank and Fox, Dieter and Hahnel, Dirk and Rosenberg, Charles and Roy, Nicholas and Schulte, Jamieson and others},
  booktitle={Proceedings 1999 IEEE International Conference on Robotics and Automation (Cat. No. 99CH36288C)},
  volume={3},
  year={1999},
  organization={IEEE}
}

@article{burgard1999experiences,
  title={Experiences with an interactive museum tour-guide robot},
  author={Burgard, Wolfram and Cremers, Armin B and Fox, Dieter and H{\"a}hnel, Dirk and Lakemeyer, Gerhard and Schulz, Dirk and Steiner, Walter and Thrun, Sebastian},
  journal={Artificial intelligence},
  volume={114},
  number={1-2},
  pages={3--55},
  year={1999},
  publisher={Elsevier}
}

@inproceedings{bouzida2024carmen,
  title={Carmen: A cognitively assistive robot for personalized neurorehabilitation at home},
  author={Bouzida, Anya and Kubota, Alyssa and Cruz-Sandoval, Dagoberto and Twamley, Elizabeth W and Riek, Laurel D},
  booktitle={Proceedings of the 2024 ACM/IEEE International Conference on Human-Robot Interaction},
  pages={55--64},
  year={2024}
}

@inproceedings{del2019lindsey,
  title={Lindsey the tour guide robot-usage patterns in a museum long-term deployment},
  author={Del Duchetto, Francesco and Baxter, Paul and Hanheide, Marc},
  booktitle={2019 28th IEEE international conference on robot and human interactive communication (RO-MAN)},
  pages={1--8},
  year={2019},
  organization={IEEE}
}

@inproceedings{sauppe2014robot,
  title={Robot deictics: How gesture and context shape referential communication},
  author={Saupp{\'e}, Allison and Mutlu, Bilge},
  booktitle={Proceedings of the 2014 ACM/IEEE international conference on Human-robot interaction},
  pages={342--349},
  year={2014}
}

@inproceedings{holladay2014legible,
  title={Legible robot pointing},
  author={Holladay, Rachel M and Dragan, Anca D and Srinivasa, Siddhartha S},
  booktitle={The 23rd IEEE International Symposium on robot and human interactive communication},
  pages={217--223},
  year={2014},
  organization={IEEE}
}

@inproceedings{huang2013modeling,
  title={Modeling and Evaluating Narrative Gestures for Humanlike Robots.},
  author={Huang, Chien-Ming and Mutlu, Bilge},
  booktitle={Robotics: Science and Systems},
  volume={2},
  year={2013}
}

@article{bartneck2008gqs,
author = {Bartneck, Christoph and Kulic, Dana and Croft, Elizabeth and Zoghbi, Susana},
year = {2008},
month = {01},
pages = {71-81},
title = {Measurement Instruments for the Anthropomorphism, Animacy, Likeability, Perceived Intelligence, and Perceived Safety of Robots},
volume = {1},
journal = {International Journal of Social Robotics},
doi = {10.1007/s12369-008-0001-3}
}

@article{obrien2018ues,
author = {O'Brien, Heather and Cairns, Paul and Hall, Mark},
year = {2018},
month = {04},
pages = {},
title = {A Practical Approach to Measuring User Engagement with the Refined User Engagement Scale (UES) and New UES Short Form},
volume = {112},
journal = {International Journal of Human-Computer Studies},
doi = {10.1016/j.ijhcs.2018.01.004}
}

@article{aeronautiques1998pddl,
  title={Pddl—the planning domain definition language},
  author={Aeronautiques, Constructions and Howe, Adele and Knoblock, Craig and McDermott, ISI Drew and Ram, Ashwin and Veloso, Manuela and Weld, Daniel and Sri, David Wilkins and Barrett, Anthony and Christianson, Dave and others},
  journal={Technical Report, Tech. Rep.},
  year={1998}
}

@article{chen2024language,
  title={Language-augmented symbolic planner for open-world task planning},
  author={Chen, Guanqi and Yang, Lei and Jia, Ruixing and Hu, Zhe and Chen, Yizhou and Zhang, Wei and Wang, Wenping and Pan, Jia},
  journal={arXiv preprint arXiv:2407.09792},
  year={2024}
}

@article{tantakoun2025llms,
  title={Llms as planning modelers: A survey for leveraging large language models to construct automated planning models},
  author={Tantakoun, Marcus and Zhu, Xiaodan and Muise, Christian},
  journal={arXiv preprint arXiv:2503.18971},
  year={2025}
}

@article{lugaresi2019mediapipe,
  title={Mediapipe: A framework for building perception pipelines},
  author={Lugaresi, Camillo and Tang, Jiuqiang and Nash, Hadon and McClanahan, Chris and Uboweja, Esha and Hays, Michael and Zhang, Fan and Chang, Chuo-Ling and Yong, Ming Guang and Lee, Juhyun and others},
  journal={arXiv preprint arXiv:1906.08172},
  year={2019}
}

@inproceedings{williams2016aggressive,
  title={Aggressive driving with model predictive path integral control},
  author={Williams, Grady and Drews, Paul and Goldfain, Brian and Rehg, James M and Theodorou, Evangelos A},
  booktitle={2016 IEEE international conference on robotics and automation (ICRA)},
  pages={1433--1440},
  year={2016},
  organization={IEEE}
}

@inproceedings{lehmann2016icubblink,
author = {Lehmann, Hagen and Roncone, Alessandro and Pattacini, Ugo and Metta, Giorgio},
year = {2016},
month = {11},
pages = {83-93},
title = {Physiologically Inspired Blinking Behavior for a Humanoid Robot},
volume = {9979},
isbn = {978-3-319-47436-6},
doi = {10.1007/978-3-319-47437-3_9}
}

@article{doughty2001consideration,
  title={Consideration of three types of spontaneous eyeblink activity in normal humans: during reading and video display terminal use, in primary gaze, and while in conversation},
  author={Doughty, Michael J},
  journal={Optometry and vision science},
  volume={78},
  number={10},
  pages={712--725},
  year={2001},
  publisher={LWW}
}

@article{gasteiger2021deploying,
  title={Deploying social robots in museum settings: A quasi-systematic review exploring purpose and acceptability},
  author={Gasteiger, Norina and Hellou, Mehdi and Ahn, Ho Seok},
  journal={International Journal of Advanced Robotic Systems},
  volume={18},
  number={6},
  pages={17298814211066740},
  year={2021},
  publisher={SAGE Publications Sage UK: London, England}
}

@inproceedings{pascher2023communicate,
  title={How to communicate robot motion intent: A scoping review},
  author={Pascher, Max and Gruenefeld, Uwe and Schneegass, Stefan and Gerken, Jens},
  booktitle={Proceedings of the 2023 CHI Conference on Human Factors in Computing Systems},
  pages={1--17},
  year={2023}
}

@inproceedings{gehle2017open,
  title={How to open an interaction between robot and museum visitor? Strategies to establish a focused encounter in HRI},
  author={Gehle, Raphaela and Pitsch, Karola and Dankert, Timo and Wrede, Sebastian},
  booktitle={Proceedings of the 2017 ACM/IEEE international conference on human-robot interaction},
  pages={187--195},
  year={2017}
}

@inproceedings{pereira2019responsive,
  title={Responsive joint attention in human-robot interaction},
  author={Pereira, Andre and Oertel, Catharine and Fermoselle, Leonor and Mendelson, Joe and Gustafson, Joakim},
  booktitle={2019 IEEE/RSJ International Conference on Intelligent Robots and Systems (IROS)},
  pages={1080--1087},
  year={2019},
  organization={IEEE}
}

@article{anzalone2015evaluating,
  title={Evaluating the engagement with social robots},
  author={Anzalone, Salvatore M and Boucenna, Sofiane and Ivaldi, Serena and Chetouani, Mohamed},
  journal={International Journal of Social Robotics},
  volume={7},
  number={4},
  pages={465--478},
  year={2015},
  publisher={Springer}
}

@inproceedings{huang2010joint,
  title={Joint Attention in Human-Robot Interaction.},
  author={Huang, Chien-Ming and Thomaz, Andrea Lockerd},
  booktitle={AAAI fall symposium: dialog with robots},
  year={2010}
}

@inproceedings{shi2024yell,
  title={Yell At Your Robot: Improving On-the-Fly from Language Corrections},
  author={Shi, Lucy Xiaoyang and Hu, Zheyuan and Zhao, Tony Z and Sharma, Archit and Pertsch, Karl and Luo, Jianlan and Levine, Sergey and Finn, Chelsea},
  booktitle={Robotics: Science and Systems},
  year={2024}
}

@inproceedings{desai2019geppetto,
  title={Geppetto: Enabling semantic design of expressive robot behaviors},
  author={Desai, Ruta and Anderson, Fraser and Matejka, Justin and Coros, Stelian and McCann, James and Fitzmaurice, George and Grossman, Tovi},
  booktitle={Proceedings of the 2019 CHI Conference on Human Factors in Computing Systems},
  pages={1--14},
  year={2019}
}

@inproceedings{unhelkar2020decision,
  title={Decision-making for bidirectional communication in sequential human-robot collaborative tasks},
  author={Unhelkar, Vaibhav V and Li, Shen and Shah, Julie A},
  booktitle={Proceedings of the 2020 ACM/IEEE International Conference on Human-Robot Interaction},
  pages={329--341},
  year={2020}
}

@article{hu2019safe,
  title={Safe navigation with human instructions in complex scenes},
  author={Hu, Zhe and Pan, Jia and Fan, Tingxiang and Yang, Ruigang and Manocha, Dinesh},
  journal={IEEE Robotics and Automation Letters},
  volume={4},
  number={2},
  pages={753--760},
  year={2019},
  publisher={IEEE}
}

@article{song2024vlm,
  title={Vlm-social-nav: Socially aware robot navigation through scoring using vision-language models},
  author={Song, Daeun and Liang, Jing and Payandeh, Amirreza and Raj, Amir Hossain and Xiao, Xuesu and Manocha, Dinesh},
  journal={IEEE Robotics and Automation Letters},
  year={2024},
  publisher={IEEE}
}

@inproceedings{li2023stargazer,
  title={Stargazer: An interactive camera robot for capturing how-to videos based on subtle instructor cues},
  author={Li, Jiannan and Sousa, Maur{\'\i}cio and Mahadevan, Karthik and Wang, Bryan and Aoyagui, Paula Akemi and Yu, Nicole and Yang, Angela and Balakrishnan, Ravin and Tang, Anthony and Grossman, Tovi},
  booktitle={Proceedings of the 2023 CHI conference on human factors in computing systems},
  pages={1--16},
  year={2023}
}

@article{bitgood2010attention,
  title={An attention-value model of museum visitors},
  author={Bitgood, Stephen},
  journal={Center for Advancement of Informal Science Education: Washington, DC, USA},
  year={2010}
}

\end{document}